\documentclass[aps,prl,twocolumn,superscriptaddress]{revtex4-1}

\usepackage[TS1,OT1,T1]{fontenc}
\usepackage{graphicx}
\usepackage{amssymb}
\usepackage{amsmath}

\newcommand{\Lagr}{\mathcal{L}}

\begin{document}
\title{Precision Metrology Meets Cosmology: Improved Constraints on Ultralight Dark Matter from Atom-Cavity Frequency Comparisons}
\author{Colin J. Kennedy}
\affiliation{JILA, National Institute of Standards and Technology and University of Colorado, Boulder, CO 80309-0440}
\author{Eric Oelker}
\affiliation{JILA, National Institute of Standards and Technology and University of Colorado, Boulder, CO 80309-0440}
\author{John M. Robinson}
\affiliation{JILA, National Institute of Standards and Technology and University of Colorado, Boulder, CO 80309-0440}
\author{Tobias Bothwell}
\affiliation{JILA, National Institute of Standards and Technology and University of Colorado, Boulder, CO 80309-0440}
\author{Dhruv Kedar}
\affiliation{JILA, National Institute of Standards and Technology and University of Colorado, Boulder, CO 80309-0440}
\author{William R. Milner}
\affiliation{JILA, National Institute of Standards and Technology and University of Colorado, Boulder, CO 80309-0440}
\author{G. Edward Marti}
\affiliation{JILA, National Institute of Standards and Technology and University of Colorado, Boulder, CO 80309-0440}
\affiliation{Department of Molecular and Cellular Physiology, Stanford University, Stanford, California 94305, United States}
\author{Andrei Derevianko}
\affiliation{Department of Physics, University of Nevada, Reno, Nevada 89557, USA}
\author{Jun Ye}
\affiliation{JILA, National Institute of Standards and Technology and University of Colorado, Boulder, CO 80309-0440}

\date{\today}

\begin{abstract}

We conduct frequency comparisons between a state-of-the-art strontium optical lattice clock, a cryogenic crystalline silicon cavity, and a hydrogen maser to set new bounds on the coupling of ultralight dark matter to Standard Model particles and fields in the mass range of $10^{-16}$ $-$ $10^{-21}$ eV. The key advantage of this two-part ratio comparison is the differential sensitivities to time variation of both the fine-structure constant and the electron mass, achieving a substantially improved limit on the moduli of ultralight dark matter, particularly at higher masses than typical atomic spectroscopic results. Furthermore, we demonstrate an extension of the search range to even higher masses by use of dynamical decoupling techniques. These results highlight the importance of using the best performing atomic clocks for fundamental physics applications as all-optical timescales are increasingly integrated with, and will eventually supplant, existing microwave timescales.

\end{abstract}

\maketitle

It is widely accepted that dark matter makes up the majority of the matter in the observable universe~\cite{schramm1998RMPbig, bertone2010BOOKparticle}. While its composition remains a mystery, its presence is manifest through gravitational effects on galactic and cosmological scales~\cite{bovy2012APJlocal,BulletCluster,Planck}. Many efforts on direct detection with laboratory experiments target weakly interacting massive particles (WIMPs) with masses in the 100 GeV range as suitable candidate particles arise naturally in popular extensions to the Standard Model. Detection has remained elusive, however, despite advances in the sensitivity for direct detection techniques~\cite{XENON1T,FERMI,cui2017dark} and missing-momentum searches at the Large Hadron Collider~\cite{aaboud2018search}.

Axions are another class of particle that has garnered interest both in solving the strong CP problem and as a dark matter candidate~\cite{safronova2018RMPsearch}. The pseudoscalar coupling of axions to matter dictates searches for odd parity signals such as permanent electric dipole moments of the neutron or electron~\cite{abel2017search,roussy2020experimental}, conversion of axions to photons in a strong magnetic field~\cite{du2018search,zhong2018results}, or nuclear recoils~\cite{aprile2020observation}. Without a clear detection of either WIMPs or axions, there is renewed interest in alternative theories proposing candidate particles with masses spanning many orders of magnitude below 1 eV down to $10^{-22}$ eV where the particle's deBroglie wavelength is comparable to the size of dwarf galaxies. This has produced new constraints arising from considerations of the astrophysical consequences of ultralight dark matter~\cite{marsh2019strong,porayko2018parkes,davoudiasl2019ultralight,irvsivc2017first}.

For scalar fields, recent efforts have been directed towards developing new laboratory probes of dark matter at particle masses far below 1 eV, as ultralight particles provide natural solutions to some astrophysical problems such as small scale structure in typical cold dark matter models~\cite{hu2000fuzzy}. In this low energy regime, the conditions that the dark matter is gravitationally bound to galaxies and it has large deBroglie wavelengths imply that these particles have large occupation numbers per mode, and consequently form a highly coherent bosonic object behaving locally like a classical field~\cite{safronova2018RMPsearch}. Specifically, ultralight bosonic dark matter with dilatonic couplings to normal matter is predicted to generate highly coherent oscillations in the fine-structure constant as well as electron and quark masses~\cite{arvanitaki2015PRDsearching}. Additionally, assuming that the dark matter contains some self-interaction, it may form clumps that can manifest as transient changes in these constants~\cite{derevianko2014NatPhyshunting, wcislo2017NatAstroexperimental, roberts2017NatureCommsearch, wcislo2018first}.

Using atomic clocks as probes of ultralight dark matter has garnered strong interest due to their inherent sensitivity to fundamental constants and the rapid advance in the clock performance over the past decade. These systems provide a promising route to advance the search frontier for ultralight dark matter particles given a wealth of historical data from global timescale systems~\cite{huntemann2014PRLimproved, godun2014PRLfrequency, blatt2008srglobal, rosenband2008Sciencefrequency, wcislo2018first} including GPS~\cite{roberts2017NatureCommsearch}, and a host of recent proposals for enhancing the sensitivity of optical lattice clocks for dark matter searches~\cite{berengut2010PRLenhanced, safronova2018PRLtwo}. 

\begin{figure}[ht]
\centering
\includegraphics[width=1.0\columnwidth]{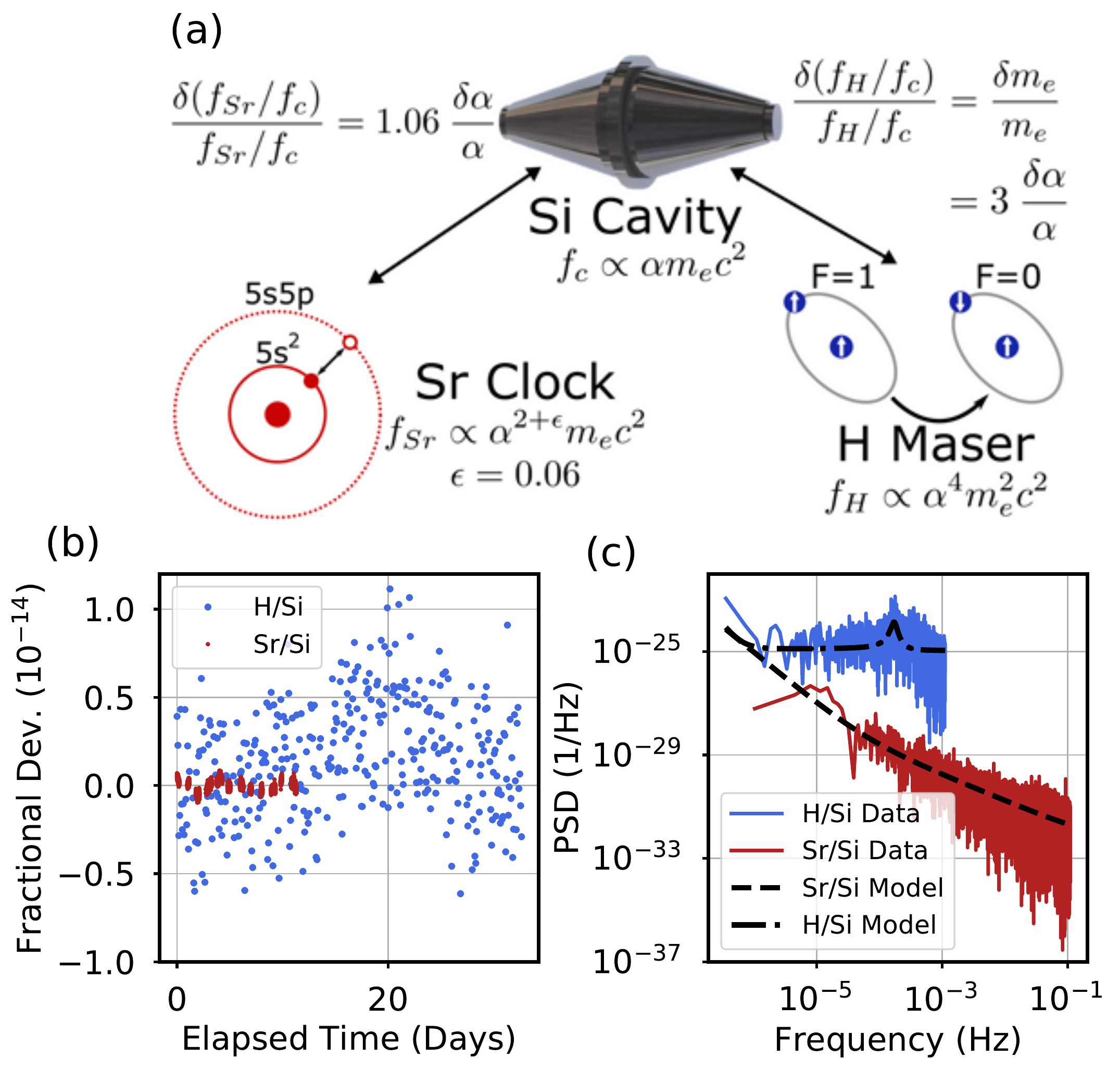}
\caption{
Atom-Cavity Frequency Comparisons. (a) Schematic depiction of the three frequencies that compose the ratios for this dark matter search. Fractional fluctuations of these ratios are sensitive to fractional fluctuations of fundamental quantities via the inscribed equations. (b) Time-domain signal of the 12-day comparison for the $f_{Sr}/f_{c}$ ratio (red) and the 33-day comparison of $f_{H}/f_{c}$ (blue). For visual clarity, $f_{H}/f_{c}$ has been decimated by a factor of 720. (c) Frequency-domain signal derived from (b) using the Lomb-Scargle technique to estimate the power spectral density (PSD). Data for $f_{Sr}/f_{c}$ has a bin width of $3.4\times10^{-6}$ Hz, and $f_{H}/f_{c}$ has a bin width of $3.7\times10^{-7}$ Hz. Black lines indicate the noise model for $f_{Sr}/f_{c}$ (dashed) and $f_{H}/f_{c}$ (dot-dashed) ratios.
}
\label{figure1}
\end{figure}

In this Letter we demonstrate significantly improved bounds on the coupling of ultralight dark matter to the fine-structure constant ($\alpha$) and electron mass ($m_e$). Central to this important advance is the state-of-the-art stability of our 21 cm crystalline silicon optical cavity at 124 K~\cite{kessler2012sub,matei20171,oelker2019demonstration} and measuring its frequency with both a Sr optical lattice clock with an estimated accuracy of $2.0\times10^{-18}$~\cite{bothwell2019} and a hydrogen maser~\cite{milner2019demonstration}. Fig.~\ref{figure1}(a) shows the cavity resonance frequency in atomic units, $f_{c} \propto \alpha m_e c^2$, and its dependence on $\alpha$ and $m_e$, both of which are directly traceable to the Bohr radius as the cavity spacer length is proportional to the silicon crystal lattice constant. The frequency of a resonant mode of the cavity is compared to atomic standards in both the optical and microwave domain forming the dimensionless frequency ratios shown in Fig.~\ref{figure1}(a). These atomic standards have different dependencies on $\alpha$ and $m_e$, making these dimensionless ratios powerful sensors for potential physics beyond the Standard Model. For the optical ratio, the $^{87}$Sr clock transition frequency depends primarily on the Rydberg energy, $\frac{1}{2}\alpha^{2+\epsilon}m_e c^2$, via the electron interaction energy difference between 5s$^{2}$ and 5s5p clock states -- with a small $\epsilon = 0.06$ correction for relativistic effects~\cite{arvanitaki2015PRDsearching}. For the microwave ratio, the $^{1}$H hyper-fine transition frequency depends on the magnetic interaction of the electron and nuclear spins, $f_{H} \propto \alpha^4 m_e^2 c^2$, leading to a different dependence on $\alpha$, and $m_e$~\cite{stadnik2016enhanced}. While not discussed in detail here, $f_{H}/f_{c}$ has additional sensitivity to the quark masses, the results of which are given in~\cite{supplement}.

Arising from this differential sensitivity to fundamental constants, precise determination of the fluctuations of both $f_{H}/f_{c}$ and $f_{Sr}/f_{c}$ provides a direct correspondence with potential fluctuations in the apparent values of fundamental constants. These relative fluctuations can subsequently be used to connect cosmological models of dark matter to atomic physics experiments via various hypotheses of how the Standard Model is coupled to dark matter. To make this connection, we consider the Lagrangian coupling terms (in natural units $\hbar = c = 1$) between dark matter, $\phi$, and Standard Model particles and fields~\cite{arvanitaki2015PRDsearching}:
\begin{equation}
    \Lagr_{\phi} = \kappa \phi \bigg( \frac{d_e}{4} F_{\mu\nu}F^{\mu\nu} - d_{m_e} m_e \Bar{\psi}_e \psi_e \bigg).
\end{equation}
The first term describes the coupling of $\phi$ to the fine-structure constant via the electromagnetic Faraday tensor, $F_{\mu\nu}$, with the electromagnetic gauge modulus $d_e$, and $\kappa = \sqrt{4\pi G_{N}}$ with Newton's constant $G_{N}$. The second term describes the coupling of $\phi$ to the electron mass with the modulus, $d_{m_e}$, and the electron field, $\psi_e$. In this model, $\Lagr_{\phi}$ produces an apparent periodic change in $\alpha$ or $m_e$, with the modulus, $d_e$ or $d_{m_e}$, respectively. This manifests in a laboratory experiment as periodic oscillations of dimensionless frequency ratios at the Compton frequency ($f_{\phi}$) which can be subsequently used to set bounds on both $d_e$ and $d_{m_e}$~\cite{supplement}.

The data composing the two dimensionless frequency ratios between the Si cavity and the Sr clock or H maser are shown in Fig.~\ref{figure1}(b). To begin, we consider a potential periodic signal that is observed over a duration of $T_{\text{max}}$, compared to the dark matter field's coherence time, $\tau_{\text{c}}$. For $T_{\text{max}} < \tau_{\text{c}}$, the phase is preserved throughout the observation run so we expect the signal-to-noise ratio (SNR) to improve as more data is collected. The $f_{Sr}/f_{c}$ data consists of 12 consecutive comparison days with a total uptime of 30\% resulting in a data set that is 978,041 s long. The $f_{H}/f_{c}$ data spans 33 days and has a total uptime of 94\% resulting in a 2,826,942 s long record. The highest frequency/mass bins for the Sr/Si data set are at 125 mHz, corresponding to a time period of 8 s, or a maximum of $1.2\times10^5$ oscillations over the course of the comparison, whereas the expected virial velocity-broadening of dark matter expects coherence times up to $10^6$ oscillations, $\tau_c = (f_{\phi} v_{\text{virial}}^2/c^2)^{-1}$~\cite{arvanitaki2015PRDsearching}. Consequently, the results of this comparison are interpreted in the limit of $T_{\text{max}} < \tau_{\text{c}}$.

The power-spectral density (PSD) of the data is shown in Fig.~\ref{figure1}(c). The $f_{Sr}/f_{c}$ model, shown overlaid in black dashed lines, is derived from previous work adding in a white frequency noise term accounting for the Dick effect~\cite{oelker2019demonstration}. The $f_{H}/f_{c}$ model, shown in dash-dotted lines, is derived from the cavity model in addition to a white frequency noise term for the maser. A resonance around 0.1 mHz accounts for the noise of the microwave link between the maser location at NIST and the cavity and optical frequency comb at JILA~\cite{supplement}. The PSD can be converted into a power spectrum using the bin width and subsequently a bound on the coupling of dark matter to Standard Model fields after accounting for the contribution of the realistic laser noise in each frequency bin.

Complications arise from the existence of time gaps in the data and the flicker frequency and random walk noise contributions in the frequency record. As a result, traditional methods for computing the power spectrum of the data are not accurate estimators of the true power spectrum. To handle the gapped data, we use a Lomb-Scargle periodogram implemented using publicly available software from the astropy Python library~\cite{vanderplas2012introduction, vanderplas2015periodograms}. In the limit of a continuous data set without gaps the Lomb-Scargle periodogram is equivalent to traditional techniques for evaluating the power spectrum even in the presence of correlated noise~\cite{supplement}. For the $f_{Sr}/f_{c}$ data the periodogram is evaluated at frequencies ranging from the sampling frequency of 125 mHz to the frequency corresponding to one oscillation over the total duration of the data set. The number of bins this frequency range is subdivided into is given by one-half of the total number of measurements taken during the observation period. Data for the $f_{H}/f_{c}$ ratio is longer in duration, has significantly fewer gaps, and is dominated by white noise, but nonetheless the power spectrum is evaluated with the same procedure as the $f_{Sr}/f_{c}$ data. Due to the worse short-term stability of $f_{H}/f_{c}$ and the presence of significant microwave link noise at high frequencies~\cite{milner2019demonstration}, data for the maser is plotted only for frequencies below 1 mHz~\cite{supplement}.

Given this technique for estimating the power spectrum of the gapped data, we next consider the amount of power observed in each frequency bin and the likelihood of observing that signal given the known laser noise model in the two-part frequency comparison. To accomplish this, the known model for each ratio is used to simulate the expected noise on a continuous data set. This data is then gapped in accordance with the timestamps of the real experiment and the Lomb-Scargle algorithm is used to estimate the power spectrum of the simulated data. This procedure is repeated 1000 times to generate a probability distribution of expected power for each frequency bin given the laser model and the existing gaps in the data~\cite{supplement}. For each bin the 95\% confidence limit can thus be determined by numerical integration of this probability distribution. This procedure is especially important for low frequency bins corresponding to timescales longer than an individual day's data as the presence of gaps in the data set reduces the sensitivity of the search to oscillations of specific frequencies. This is seen most prominently at masses ranging from $4\times10^{-20}$ to $1\times10^{-18}$ eV arising from the diurnal pattern of data collection, with more discussion on data gaps provided in~\cite{supplement}.

\begin{figure}[ht]
\centering
\includegraphics[width=1.0\columnwidth]{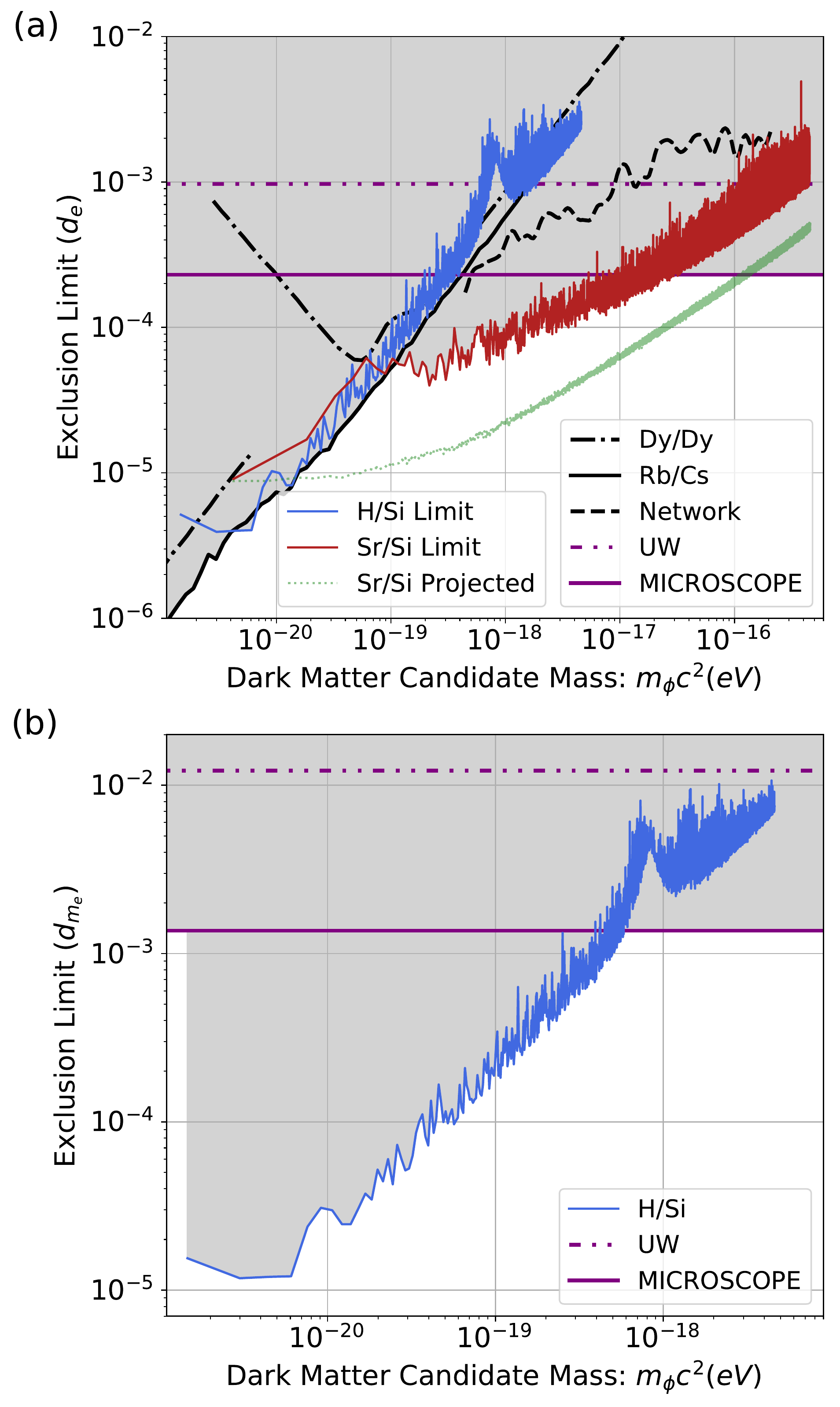}
\caption{
Dilatonic Dark Matter Exclusion Plot. (a) 95\% confidence limits placed on the electromagnetic gauge modulus, $d_{e}$, showing the improved limits set by $f_{Sr}/f_{c}$ (red) and $f_{H}/f_{c}$ (blue) ratios in the mass range above $1\times10^{-19}$ eV. The maximum projected sensitivity for a search of the same 11-day duration without data gaps is included (light green) highlighting the potential of this technique. Limits derived from previous spectroscopic searches (black lines)~\cite{vanTilberg2015PRLsearch, hees2016PRLsearching, wcislo2018first} and equivalence principle tests (purple lines)~\cite{schlamminger2008PRLtest, berge2018PRLmicroscope} are included. (b) Demonstration of a significantly improved limit on the electron mass modulus ($d_{m_e}$) derived from the $f_{H}/f_{c}$ (blue) ratio. Limits from equivalence principle tests (purple lines)~\cite{schlamminger2008PRLtest, berge2018PRLmicroscope} are included. Shaded regions in both (a) and (b) are excluded at the 95\% confidence level given the observed signal and noise models.
}
\label{figure2}
\end{figure}

The amplitude spectrum is then derived from the observed and expected power spectrum and the direct relation between the fractional amplitude in each bin and limits on $d_e$ and $d_{m_e}$ is determined. To provide a rigorous confirmation of all limits derived in this paper, we inject a periodic signal into the analysis pipeline with a known frequency and amplitude to demonstrate that our approach provides the correct sensitivity limit for detection~\cite{supplement}. In the analysis we subtract the overall linear drift of the laser cavity or the hydrogen maser. For this reason we do not consider any frequencies below the inverse duration of the data set. With artificial signal injection, the impact of this subtraction only underestimates the lowest frequency bin by a factor of 2.3 and is corrected accordingly.

Fig.~\ref{figure2}(a) shows the limits on $d_e$ derived from both $f_{Sr}/f_{c}$ (red) and $f_{H}/f_{c}$ (blue). These results surpass previous limits, and particularly at higher masses, on $d_e$ set by atomic spectroscopy (black lines) of Dy~\cite{vanTilberg2015PRLsearch}, Rb and Cs atomic clocks~\cite{hees2016PRLsearching}, and an optical clock network~\cite{wcislo2018first}. Fig.~\ref{figure2}(b) shows the limits on $d_{m_e}$ derived from $f_{H}/f_{c}$ (blue). 

As recently highlighted in Ref.~\cite{centers2019stochastic}, interferences of different amplitudes that are thought to compose the local thermal state of the dark mater field create a distribution of possible field amplitudes, $\phi$, observed locally. As the apparent variation of constants comes from $\phi$ acting on $\alpha$ or $m_e$, the limit for the moduli $d_e$ and $d_{m_e}$ thus depends on the value of $\phi$ experienced locally. When $T_{\text{max}} < \tau_{\text{c}}$, the experiment is sampling from this distribution once and the non-deterministic amplitude affects the exclusion region of $d_e$ and $d_{m_e}$. This effect is captured in Fig.~\ref{figure2} by a uniform rescaling of all spectroscopic limits previously published by a factor of 3.0 in accordance with the 95\% limit of this distribution.

At the same time, limits on $d_e$ and $d_{m_e}$ are also set by equivalence principle tests~\cite{kuroda1989PRLtest, williams2012CQGlunar, berge2018PRLmicroscope, schlamminger2008PRLtest} with the most stringent limits to date coming from measurements of differential accelerations of macroscopic masses~\cite{berge2018PRLmicroscope, schlamminger2008PRLtest}, plotted in Fig.~\ref{figure2} (purple lines). These limits are not affected by the stochastic amplitude of $\phi$ because the coupling to dark matter in this case arises through a virtual exchange of the ultralight dark matter particle that mediates a Yukawa potential. Therefore, these searches are not affected by the rescaling which spectroscopic searches undergo.

Fig.~\ref{figure2} shows the limits established by both $f_{Sr}/f_{c}$ and $f_{H}/f_{c}$ on $d_e$ and $d_{m_e}$ at the 95\% confidence level. For $d_e$, we set a new bound at a range of masses from $1\times10^{-17}$ to $1\times10^{-19}$ eV, and improve upon the limits coming from atomic spectroscopy (black lines) from $4.5\times10^{-16}$ down to $1\times10^{-19}$ eV, improving the limit by a factor of five in certain mass ranges. One prominent, high-$Q$ peak at $3.83\times10^{-16}$ is identified as arising from magnetic field noise which is imperfectly rejected by the atomic servo. For $d_{m_e}$, $f_{H}/f_{c}$ is the only data with sensitivity, and here we improve the limit by up to a factor of 100 between $2\times10^{-19}$ and $2\times10^{-21}$ eV. This disfavors $d_{m_{e}}$ at the level of $\sim4\times10^{-5}$ for the lowest masses, and can easily be improved and extended to lower candidate masses when longer data sets are collected. For comparison, Ref.~\cite{arvanitaki2016sound} discusses natural values of both $d_e$ and $d_{m_e}$, and shows that the current limits on $d_{m_e}$ are $\sim 10^{6}$ above natural values for this parameter. In total, the data does not support a signal consistent with coupling of dark matter to standard model particles and fields subject to the constraints of the known cavity and maser noise models.

\begin{figure}[ht]
\centering
\includegraphics[width=1.0\columnwidth]{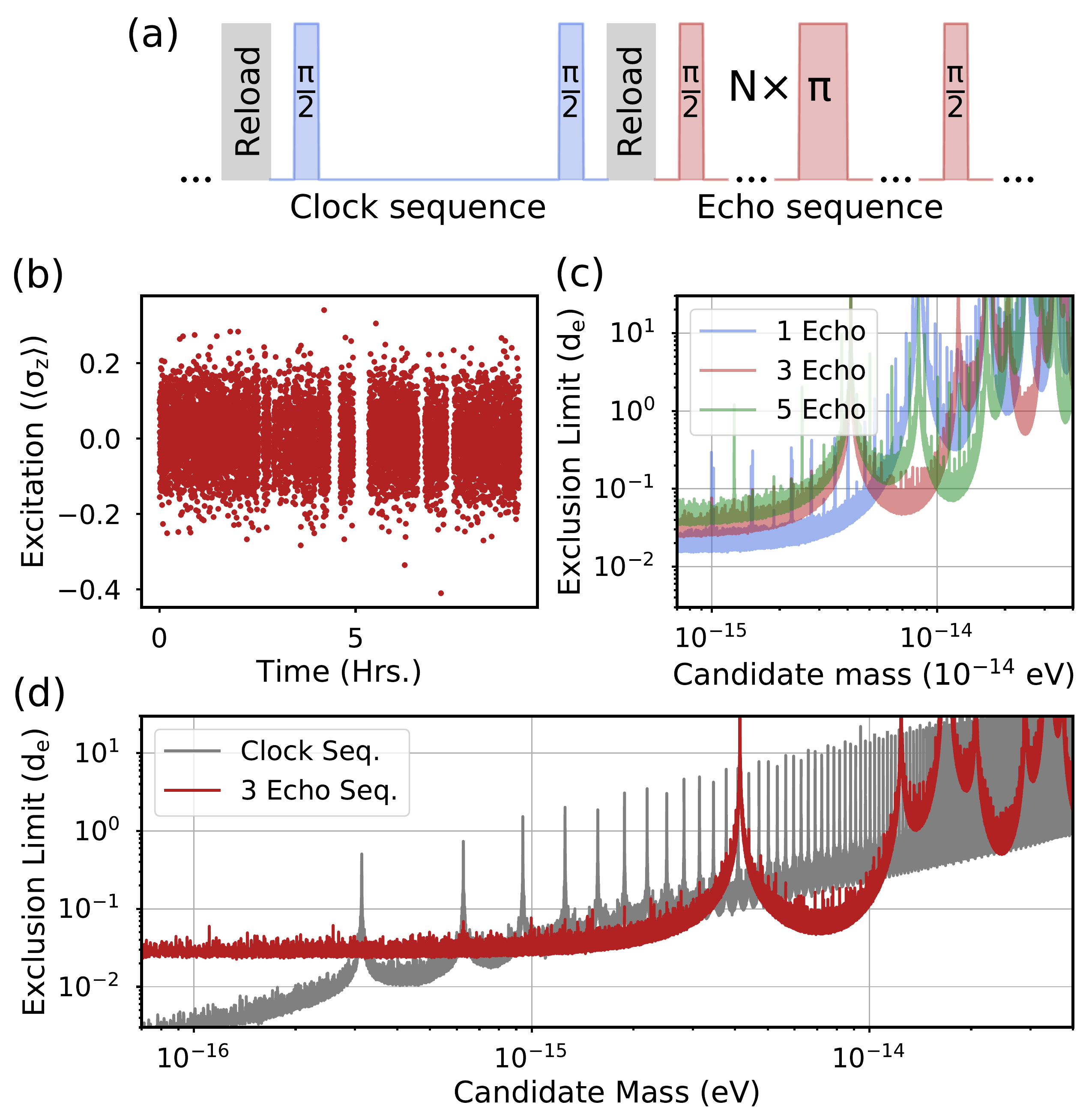}
\caption{High Frequency Search with Dynamical Decoupling. (a) Experimental sequence to interleave clock and spin-echo pulses. The dark time for both sequences is 1 second, with the number of echo pulses, N = 1,3, and 5. (b) Time-domain signal for the three-pulse sequence over single day of data collection, or 33,043 seconds. (c) Limits derived from each of the three pulse sequences demonstrating the flexibility in tailoring the sensitivity function for targeted searches. All resonances shown are consistent with the known laser model. (d) Comparison of the limit derived from interpolating the low-frequency data of the clock sequence to higher frequencies (grey) compared to the dynamical decoupling sequence for N=3 (red) showing clear advantage for frequencies above the Nyquist frequency and the customized sensitivity for specific mass regions.}
\label{figure3}
\end{figure}

To expand the sensitivity of the search to a higher mass range beyond $10^{-16}$ eV~\cite{geraci2019}, we demonstrate a technique of dynamical decoupling during the operation of the Sr lattice clock~\cite{bishof2013dynamic,carr1954effects,aharony2019constraining,kolkowitz2016gravitational}. Figure~\ref{figure3}(a) details the time sequence for this search, consisting of interleaved Ramsey and echo sequences used for clock-operation and high-frequency sampling, respectively. The clock sequence locks the laser to the atomic resonance, and enables near-continuous probing of the high frequency noise for many consecutive hours. Fig.~\ref{figure3}(b) shows the resulting measurements of the excitation fraction for the case of a three-echo pulse sequence (N=3) converted into units of $\langle \sigma_{z} \rangle$, in the basis of \{$\vert g \rangle$, $\vert e \rangle$\} ground- and excited-clock states. For a highly coherent oscillation, extending the total observation time up to the expected coherence time of the dark matter field increases the search sensitivity. Here, each echo sequence was run for one working day, limiting the data length and thus the ultimate limit that can be set by this technique. With longer observation times, an expected improvement of $\sim$100 is attainable thereby making such a search compelling.

For the high-frequency search, we conduct three separate experiments varying the number of echo pulses between N=1, 3, and 5. For all pulse sequences, including the clock sequence, we use an evolution time of 1 s between the initial and final $\pi/2$-pulses with $\pi$-pulses of 28 ms in duration. Each search produces two signals for $f_{Sr}/f_{c}$, one at low frequencies coming from the control and error signals of the clock sequence and another at high frequencies coming from the echo sequence. Each is converted to a limit at masses above the sampling frequency of the clock using the analysis outlined in~\cite{supplement}. Fig.~\ref{figure3}(c) shows the resulting limits for each of the three dynamical decoupling pulse sequences. The ability to tailor the search range and sensitivity is apparent. Additionally, many sharp resonances appear in the spectrum which are found to be consistent with the known laser model by comparison with simulation.

Fig.~\ref{figure3}(d) demonstrates the relative advantage of dynamical decoupling when compared to an extension of a low-frequency search to higher frequency. Here we see two main advantages. First, the ability to tune the point of maximal sensitivity is evident by the lower limit achieved at masses around $7\times10^{-15}$ eV. Second, the limit derived from dynamical decoupling does not suffer from reduced sensitivity at every multiple of the Nyquist frequency as is apparent in the limit derived from the clock sequence. Note that the exclusion limits shown in Fig.~\ref{figure3}(c,d) have been corrected for the assumption of a stochastic dark matter field amplitude as in Fig.~\ref{figure2}.

The new limits demonstrated in this work highlight how the search for new physics is enabled directly by the higher degree of frequency stability demonstrated by cryogenic Si cavities~\cite{oelker2019demonstration} as well as the improved accuracy, stability, and long-term reliability of the optical lattice clock~\cite{bothwell2019}. This improved stability allows observing the fractional variation of the frequency ratios in this three-element network with a higher degree of precision than previously possible. This also shows that as the stability of optical cavities and the precision and total uptime of optical clocks improve over time, each advance enables the re-examination of this result with progressively higher resolution. Integrated with a microwave timescale, future timekeeping systems will be able to extend the discovery reach of these searches for dark matter induced variation of fundamental constants. Indeed, the data presented here is a subset of that used to demonstrate an all-optical timescale with record low timing error over one month of operation~\cite{milner2019demonstration}. As ultrastable laser technologies and laser cooling techniques advance to include atomic species with highly relativistic clock transitions in neutral atoms, ions, highly charged ions, and nuclei, the resolution with which these effects can be resolved is expected to advance greatly. 

We thank T. Fortier, J. A. Sherman, and H. Leopardi for the maser connection, D. Matei, T. Legero, U. Sterr, and F. Riehle for the collaboration on silicon cavities, and S. Campbell, R. Hutson, A. Goban, S. Kolkowitz, Y. Stadnik, K. Van Tilburg, S. Dimopoulos, and M. Arvanitaki for useful discussions. We acknowledge funding support from NIST, DARPA, NSF Phys-1734006, and PHY-1912465.


%

\newpage
\renewcommand\thefigure{S\arabic{figure}}
\setcounter{figure}{0}

\section{Supplementary Material}

\subsection{Quark mass limits}

\begin{figure}[ht]
\centering
\includegraphics[width=1.0\columnwidth]{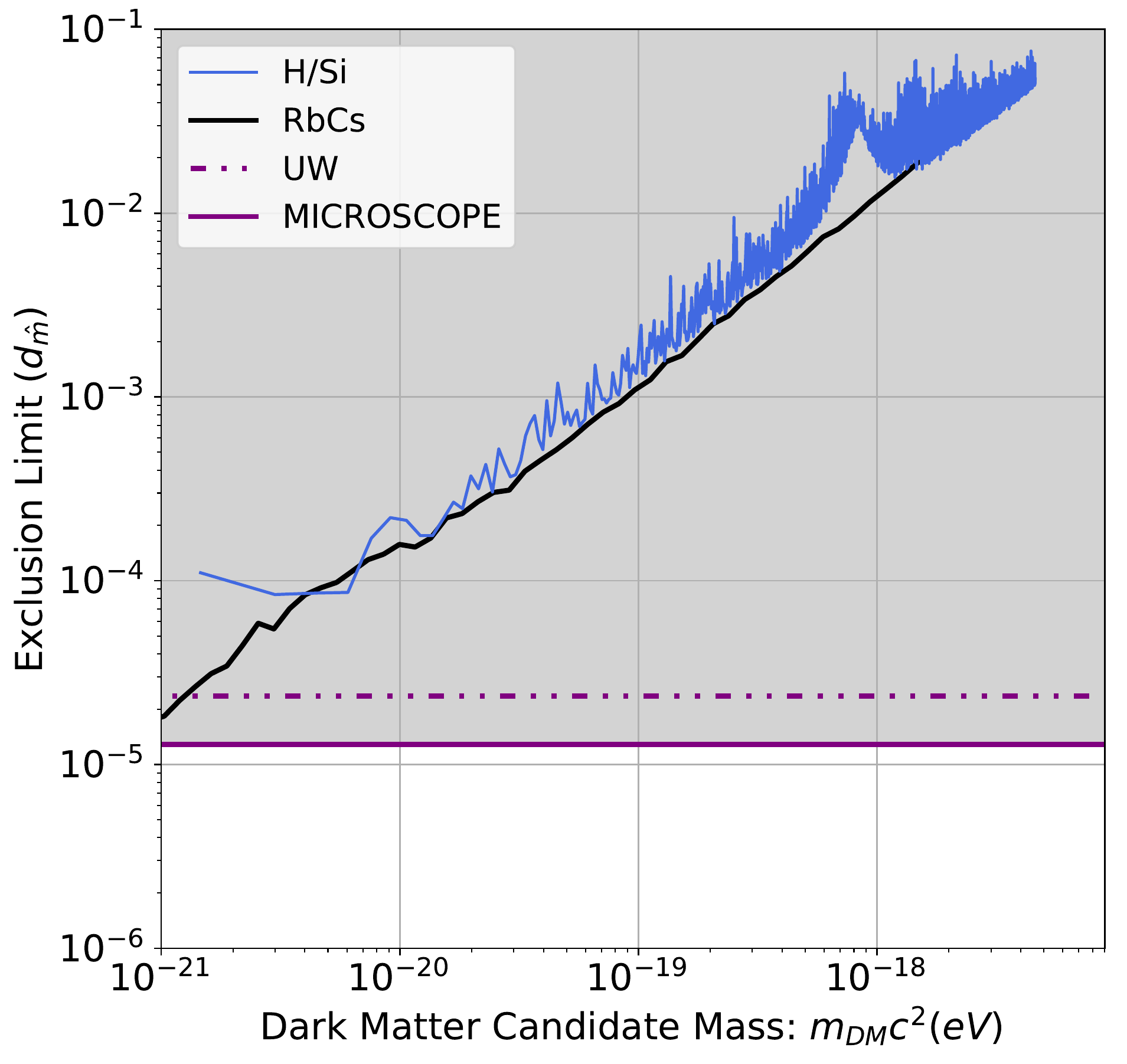}
\caption{Quark Mass Coupling Limits. Fluctuations of the $f_{H}/f_{c}$ ratio has a third sensitivity in addition to the fine-structure constant and electron mass presented in the main text which is sensitivity to the modulus of the average quark mass, $d_{\hat{m}}$. Analysis of this ratio allows a bound to be set on $d_{\hat{m}}$ that, although it does not set a new limit, is complimentary to existing limits given the different experimental platforms used to derive this limit compared to Refs.~\cite{schlamminger2008PRLtest, berge2018PRLmicroscope, hees2016PRLsearching}.}
\label{figure_s1}
\end{figure}

In addition to the sensitivity of the $f_{H}/f_{c}$ ratio to fluctuations of the fine-structure constant and the electron mass, there is an additional sensitivity to the modulus of the average up- and down-quark mass, $d_{\hat{m}}$. Fig. \ref{figure_s1} shows the limit derived from the  $f_{H}/f_{c}$ ratio for  $d_{\hat{m}}$. Although it is not competitive with the state-of-the-art limits derived from equivalence principle tests, the limit provides a complimentary check on these results in a radically different experimental platform, and reinforces the limits derived from atomic spectroscopy by the comparison of atomic clocks in Rb and Cs~\cite{hees2016PRLsearching}. As such, this result provides a useful cross-check of the results offered from Refs.~\cite{berge2018PRLmicroscope, hees2016PRLsearching, schlamminger2008PRLtest}.

\subsection{Dynamical Decoupling Sensitivity}

\begin{figure}[ht]
\centering
\includegraphics[width=1.0\columnwidth]{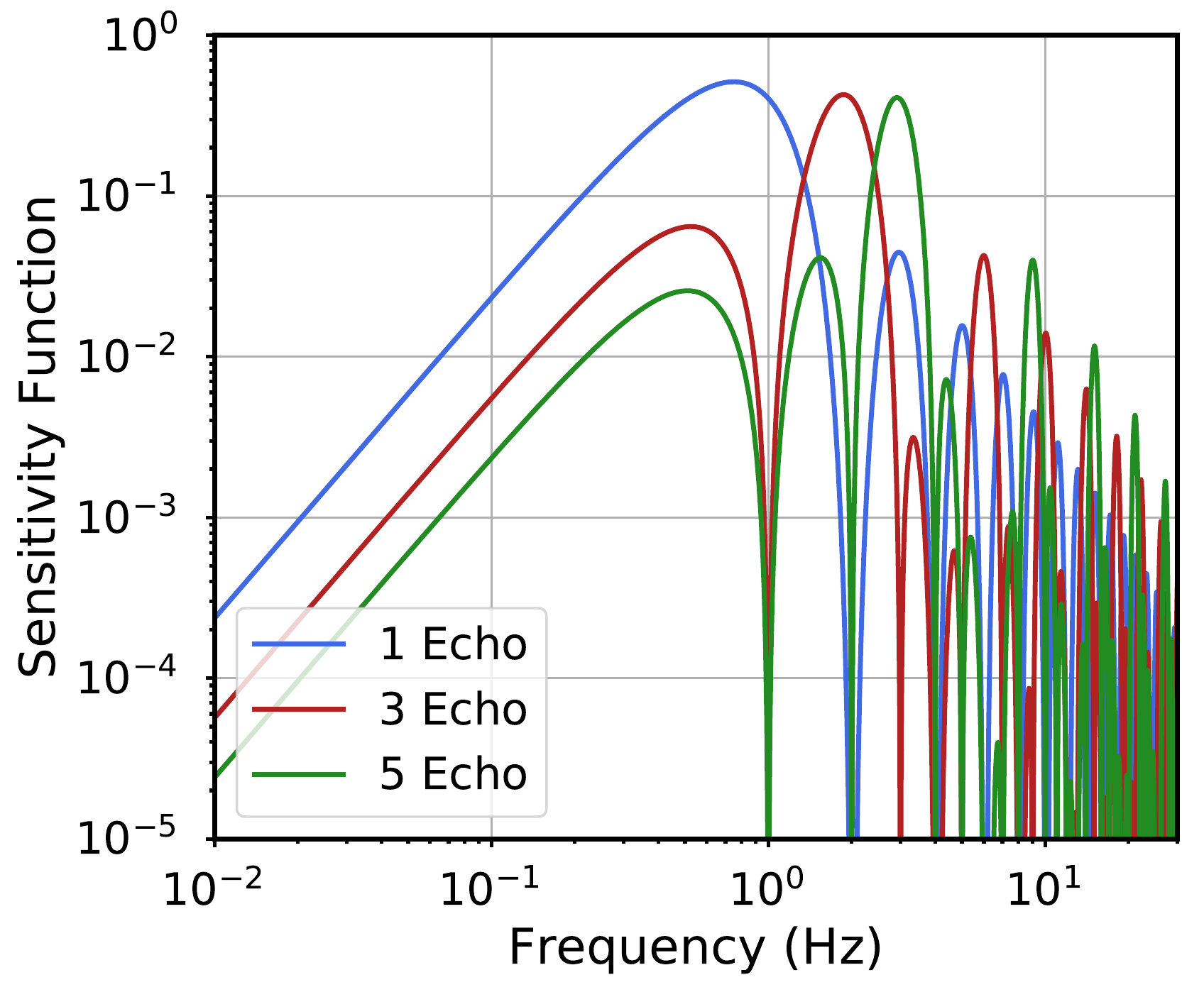}
\caption{Sensitivity functions for the dynamical decoupling $d_e$ limits. Functions for 1 (blue), 3 (red), and 5 (green) echo pulses determine how noise at each frequency is downsampled into the base band and conversely how the observed power in the base band translates to a limit at higher frequencies.}
\label{figure_s2}
\end{figure}

One main advantage of a dark matter search with dynamical decoupling sequences featuring different numbers of echo-pulses and thus differing sensitivities is the ability to tune the sensitivity of a search to a specific mass range with greater sensitivity than what is allowed by searching for aliased high-frequency signals in a Ramsey or Rabi sequence. Figure \ref{figure_s2} demonstrates this tunable sensitivity in the three limits for $d_e$ derived from this search resulting from different pulse sequences using 1, 3, and 5 echo-pulses. The functions plotted here are for a total dark time between initialization and readout pulses of 1 second, and a $\pi$-pulse time of 28 ms. 

The zeros in the sensitivity function correspond to regions in the main text Fig.~\ref{figure3}(c) where the derived limit has broad peaks corresponding directly to this loss of sensitivity. The sharp peaks which appear in the spectrum in Fig.~\ref{figure3}(c) come from known resonances in the laser model and can be verified by comparison of simulated data and the collected data. The absence of a large peak in the N=3 data suggests that the 5.7 Hz peak in the laser model of Ref.~\cite{oelker2019demonstration} is not observed in this work and is subsequently omitted from the $f_{Sr}/f_{c}$ model. Note that without varying the pulse sequence timing this method cannot uniquely identify the frequency from which the baseband signal is aliased from due to the many-to-one mapping of frequency noise on a given experimental cycle to fluctuations in the polar angle of the collective spin vector. As a result, we set bounds on the amplitude of a peak at every potential high-frequency component which may be aliased down to a given baseband frequency. This results in the series of peaks repeated throughout the limits in Fig.~\ref{figure3}(c).

\subsection{Derivation of Limits from Power Spectra}

Following the formalism laid out in Ref.~\cite{vanTilberg2015PRLsearch}, the fractional variation of a fundamental constant induced by $\mathcal{L}_{\phi}$ can be expressed as:
\begin{eqnarray}
    \frac{\delta \alpha(t)}{\alpha} = d_{e} \kappa \phi(t), \\
    \frac{\delta m_{e}(t)}{m_{e}} = d_{m_{e}} \kappa \phi(t).
\end{eqnarray}

\begin{figure}[ht]
\centering
\includegraphics[width=1.0\columnwidth]{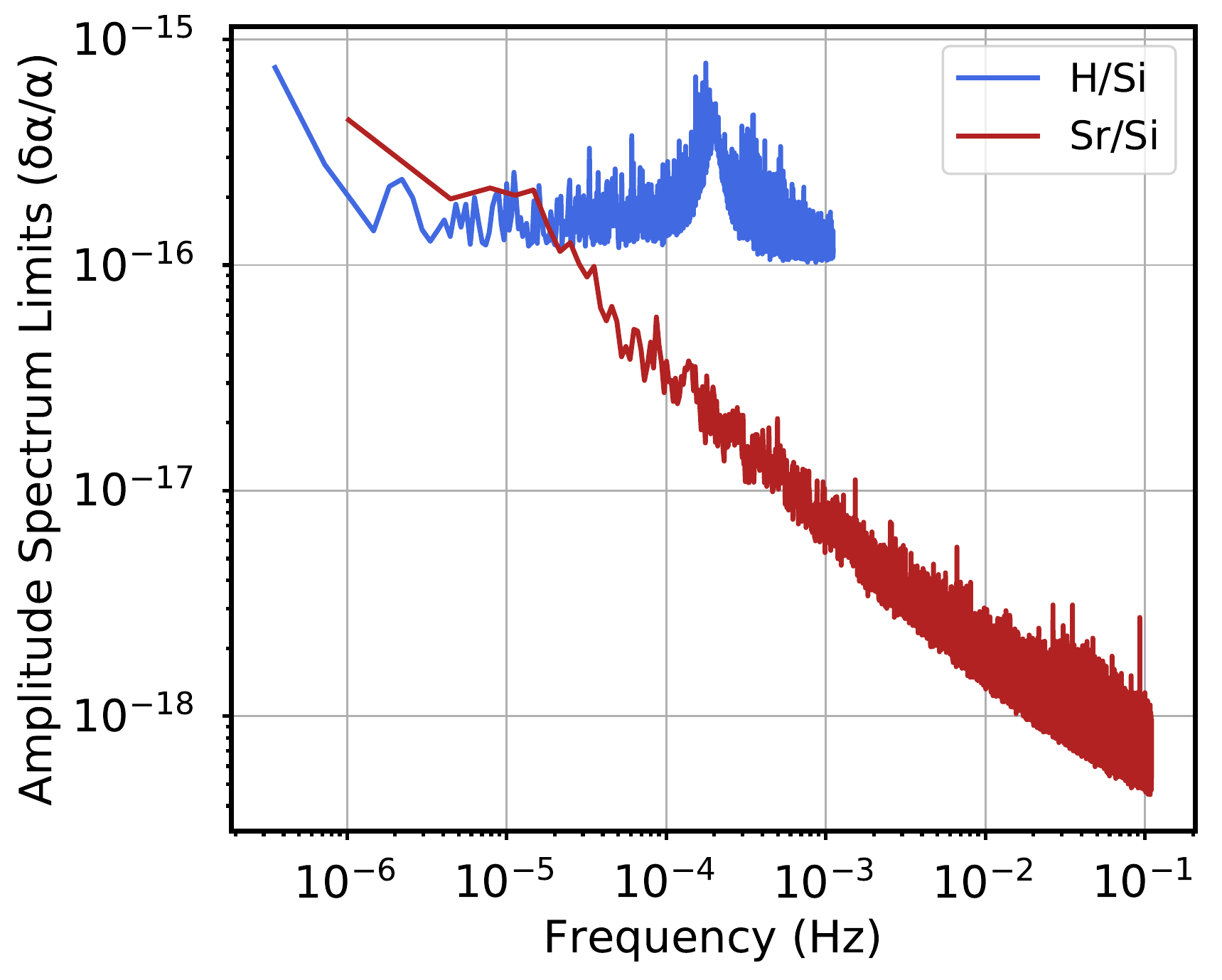}
\caption{
Amplitude Spectrum Confidence Limits. 95\% confidence limits on the amplitude of $\alpha$ variation from both $f_{H}/f_{c}$ (blue) and $f_{Sr}/f_{c}$ (red).
}
\label{figure_s2_5}
\end{figure}

For example, the confidence limit on $\alpha$ variation is shown in the amplitude spectrum in Fig.~\ref{figure_s2_5}.
Without applying any specific model to interpret either the $f_{H}/f_{c}$ or $f_{Sr}/f_{c}$ ratio, we can already use the data in Fig.~\ref{figure_s2_5} to bound periodic fluctuations in $\alpha$ arising from all sources. On the other hand, any particular theoretical model which produces periodic oscillations in $\alpha$ may use this limit as a starting point to produce a bound on a microscopic parameter such as a coupling constant.

In this case, the periodic form of the classical wave solution that describes the dark matter field, $\phi$ can be used along with the relations between fractional frequency fluctuations between atom $A$, $f_A$, and the cavity frequency, $f_c$, and fractional fluctuations in the fundamental constants to write:
\begin{eqnarray}
    \frac{\delta(f_A/f_c)}{f_A/f_c} = K_{\text{A,c}} d_e \kappa \phi(t), \\
    \frac{\delta(f_A/f_c)}{f_A/f_c} = K_{\text{A,c}} d_{m_e} \kappa \phi(t),
\end{eqnarray}
where in each case $K_{\text{A,c}}$ is the specific sensitivity of that frequency ratio to that variation. To turn this into a limit on the moduli, $d_e$ and $d_{m_e}$, we must make some key assumptions about the form of $\phi(t)$. The first assumption, discussed in Ref.~\cite{centers2019stochastic}, is that the local amplitude of $\phi(t)$ is sampled from a distribution of possible values given the stochastic nature of the field. As a result, the amplitude of the time-dependent field, $\phi_0$ is treated as a random variable with a known distribution. Next, we also assume that $\phi$ accounts for the entirety of the conservatively estimated 0.3 GeV/cm$^{3}$ local dark matter energy density such that:
\begin{equation}
    \kappa \phi_0 = 6.4\times10^{-13} \Bigg( \frac{10^{-18} \text{eV}}{m_{\phi}} \Bigg).
\end{equation}
Next, the amplitude spectrum of fractional ratio fluctuations is calculated from the data by, $\sqrt{S_y(f)}$, where $S_y(f)$ is the power spectrum in the unit-less fractional ratio at frequencies $f$. Frequencies in units of Hz are converted to mass units in multiplying by $4.14\times10^{-15}$ eV/Hz. After some rearrangement, the limit on the modulus at each mass bin in the amplitude spectrum is given by:
\begin{equation}
    d_e = \xi\frac{\sqrt{S_y(f)}}{K_{\text{A,c}}}\:\frac{f}{1.55\times10^{-16}}
\end{equation}
where $K_{\text{A,c}}$ is the sensitivity for the particular frequency ratio (1.06 for Sr/Si, 3 for H/Si). From Ref.~\cite{centers2019stochastic}, this limit applies to one particular value of $\phi_0$ sampled from the possible distribution of values, so to account for this effect, the limit is scaled by the 95\% confidence limit of this distribution given by $\xi$. Conversion of the observed data to electron mass limits proceeds in an identical manner but with a different sensitivity coefficient. 

The dynamical decoupling sequence uses a series of $\pi$ echo pulses to tailor the sensitivity function of the atomic response to probe fluctuations of the atom-cavity detuning at frequencies higher than the sampling frequency. For this sequence, the conversion from an observed excitation fraction variation to a exclusion plot includes an extra step to account for the signal being one that is aliased from higher frequencies down to the baseband where it is observed. Since the noise is dominated by dephasing from the local oscillator, the noise power in both quadratures is equivalent and pulses $\pi/2$ rotated from the local oscillator as implemented in alternate dynamical decoupling schemes~\cite{biercuk2009experimental} are not implemented here. We start with the polar angle of a classical spin vector, $\theta_n$ resulting from the $n$-th experimental cycle, and time-dependent atom-cavity detuning $\Delta(t')$ given by:
\begin{equation}
    \theta_n = \int_{-\infty}^{\infty} dt' \: \Delta(t')r(t'- nT_c),
\end{equation}
where $r(t'-nT_c)$ is the time-domain sensitivity function for the $n$-th experiment with cycle time $T_c$, and the polar angle is related to the excitation fraction, $p_e$, by $p_e = \sin^2{\theta/2}$. In the experiment, we collect a long data set of many repeated samples of $\theta_n$, and compute the power spectral density of the observed data set. Using the Wiener-Khinchin theorem, this power spectrum can be expressed as the Fourier transform of the autocorrelation function,
\begin{equation}
    S_{\theta}(f) = \frac{1}{N} \sum_{m=1}^{N} R_{\theta}(m) e^{-2\pi if mT_c},
\end{equation}
where $R_{\theta}(m) = \langle \theta_n \theta_{n+m} \rangle$ is the autocorrelation of the long data set of $\theta_n$ measurements.

Using the given relation between the time-dependent atom-cavity detuning and the measured polar angle we can rewrite the autocorrelation in $\theta_n$ as one in $\Delta$ by,
\begin{equation}
    \iint dt_1 \, dt_2 \, \langle \Delta(t_1) \Delta(t_2)\rangle r(t_1 - nT_c) r(t_2 - nT_c - mT_c),
\end{equation}
The sensitivity function $r(t_1 - nT_c)$ corresponds to the single-cycle sensitivity function time-shifted to the $n$-th experimental cycle. However, for a stationary process we can set $n = 0$ without loss of generality. Similarly, this means that the autocorrelation in detuning only depends on the time difference, $R_{\Delta}(t_2-t_1)$, so the integral is simplified to:
\begin{equation}
    \iint dt_1 \, dt_2 \, R_{\Delta}(t_2-t_1) r(t_1) r(t_2 - mT_c).
\end{equation}

We aim to simplify this complicated time integration by moving to the frequency domain via Parseval's theorem. Note that $r(t)$ and $\Delta(t)$ are purely real functions, but this process can we well-defined for complex functions as well. We can now clearly identify the integration over $t_1$ as the convolution of the detuning autocorrelation function and the sensitivity function, $r(t_1)$, leaving just the integration over $t_2$:
\begin{equation}
    \int_{-\infty}^{\infty} dt_2 \: (R_{\Delta}*r)(t_2) r(t_2-mT_c),
\end{equation}
where we denote the convolution with the common $(A*B)(t) = \int d\tau \: A(t-\tau)B(\tau)$ syntax. Next, we use Parseval's theorem to move the integration of the time domain signal to the integration of the frequency domain signal. Denoting the Fourier transform as $F[...]$, the integral is equivalent to:
\begin{equation}
    \int_{-\infty}^{\infty} df' \: F[(R_{\Delta}*r)(t_2)] F^{*}[r(t_2-mT_c)],
\end{equation}
from which we define the frequency domain representation of the sensitivity function, $R(f) = F[r(t)]$. Note that here the phase of the intervening $\pi$ pulses in the echo sequence are aligned with the beginning Ramsey pulse.

We can now use the convolution theorem to write, $F[(R_{\Delta}*r)] = F[R_{\Delta}(t)]F[r(t)]$, and also use the Fourier shift theorem to write,
\begin{equation}
    F^{*}[r(t_2-mT_c)] = e^{2\pi if'mT_c}F^{*}[r(t_2)].
\end{equation}
Finally, we again use the Wiener-Khinchin theorem to write the Fourier transform of the autocorrealtion function as the power spectral density, $S_{\Delta}(f) = F[R_{\Delta}(t)]$. As a result, we can express the power spectral density of polar angle fluctuations that we observe as,
\begin{equation}
    S_{\theta}(f) = \frac{1}{N}\sum_{m=1}^{N} \int_{-\infty}^{\infty} df' \: S_{\Delta}(f') \vert R(f') \vert^2 e^{-2\pi i (f-f') mT_c},
\end{equation}
where $S_{\Delta}(f')$ is the power spectrum of the relative atom-cavity detuning fluctuations which the dark matter signal acts on, and $\vert R(f') \vert^2$ is known from the pulse sequence and is plotted in the previous section. 

The remaining sum can be straightforwardly evaluated, now that the argument of the integral has lost all $m$-dependence, and is immediately identified as a Kronecker delta function, $\delta_{f',f}$. This leads to the expression of the observed power spectrum, $S_{\theta}(f)$ in terms of the known sensitivity function, $\vert R(f)\vert^2$ and the power spectrum of high frequency atom-cavity frequency fluctuations that we wish to infer, $S_{\nu}(f)$, as:
\begin{equation}
    S_{\theta}(f) = S_{\Delta}(f) \vert R(f) \vert^2.
\end{equation}
Finally, we arrive at an expression for the inferred power spectrum written in fractional frequency units of the $^{87}$Sr clock frequency, $f_{Sr}$:
\begin{equation}
    S_{\nu}(f) = \frac{S_{\theta}(f)}{(2\pi f_{Sr})^2 \vert R(f) \vert^2}.
\end{equation}

Using this expression, the estimate of the power spectrum at frequencies above the Nyquist frequency can be established by evaluating the Lomb-Scargle periodogram at each frequency and then dividing this spectrum by the appropriate sensitivity function to reflect how sensitive the power observed in the baseband is to fluctuations at a given high frequency. Numerical simulation of injected signals at high frequencies have additionally been used to verify the analysis pipeline and to ensure proper normalization of the resulting limit.

\subsection{Effect of Data Gaps on the Lomb-Scargle PSD}

\begin{figure}[ht]
\centering
\includegraphics[width=1.0\columnwidth]{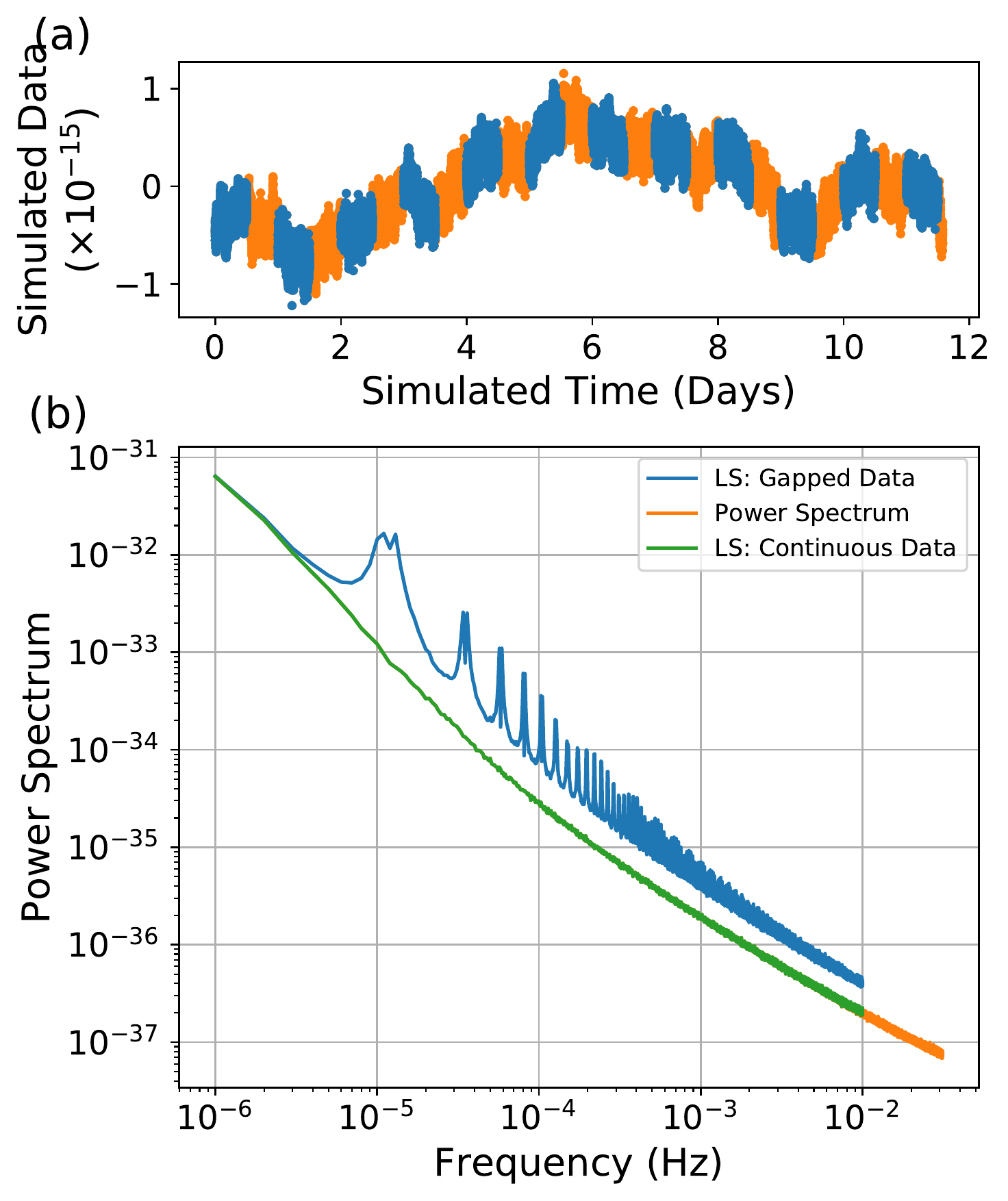}
\caption{Effect of Gapped Data. (a) Simulation of realistic $f_{Sr}/f_{c}$ data with a continuous time record (orange) which is to be compared to the identical data which has been cut into a repeating pattern of 12 hours of continuous measurement followed by 12 hours of measurement dead-time (blue). (b) The resulting power spectra of 1000 independent simulations averaged for visual clarity. The power spectrum of the continuous data set (orange) is shown compared to the Lomb-Scargle evaluation of the same data (green) and the Lomb-Scargle evaluation of the data set with 50\% uptime in 12-hour intervals (blue), showing the nontrivial effect that gaps in the data set have in the estimation of the power spectrum of gapped data.}
\label{figure_s3}
\end{figure}

Due to the finite uptime of realistic elements of current and future timescales, the data sets used in the main text have gaps in the time record. To handle these gaps, we utilized an algorithm deploying the Lomb-Scargle periodogram to estimate the power spectrum of this data. To understand the impact of the gaps in the time record, we consider the simulated experiment, shown in Fig. \ref{figure_s3}(a), where over the course of $1\times10^{6}$ seconds ($\sim$12 days) the $f_{Sr}/f_{c}$ ratio is sampled for the same 12 hours of a 24-hour period over the course of the observation run. From this simulated experiment, the power spectrum of the simulated data can be computed using both traditional means as well as using the Lomb-Scargle technique, showing the exact agreement of the two shown in Fig. \ref{figure_s3}(b). The highest frequencies of the Lomb-Scargle periodogram are omitted for clarity so that the overlap with the power spectrum can be shown. However, when gaps are added to the same data set by removing 12 hours of every 24-hour period, significant degradation in the accuracy of the Lomb-Scargle technique is apparent. Three main effects are noted. First, due to the fact that there is only half as much data in the gapped data set, the power at high frequencies is higher by a factor of two. Next, prominent peaks at odd harmonics of the frequency corresponding to the time between data gaps, $1.16\times10^{-5}$ Hz, appear in the spectrum. The loss of sensitivity to power at these frequencies can be understood by considering a sine wave with frequency and phase set such that the wave's crest occurs exactly synced up with observation periods. Finally, for frequency bins corresponding to times longer than a single day, the estimated power agrees with that observed for a continuous data set. The observed differences between the gapped and the continuous time record decreases as the amount of dead time is decreased, highlighting both the importance of high uptimes of the frequency network in a dark matter search, as well as the elimination of large gaps in the data record. 

\begin{figure}[ht]
\centering
\includegraphics[width=1.0\columnwidth]{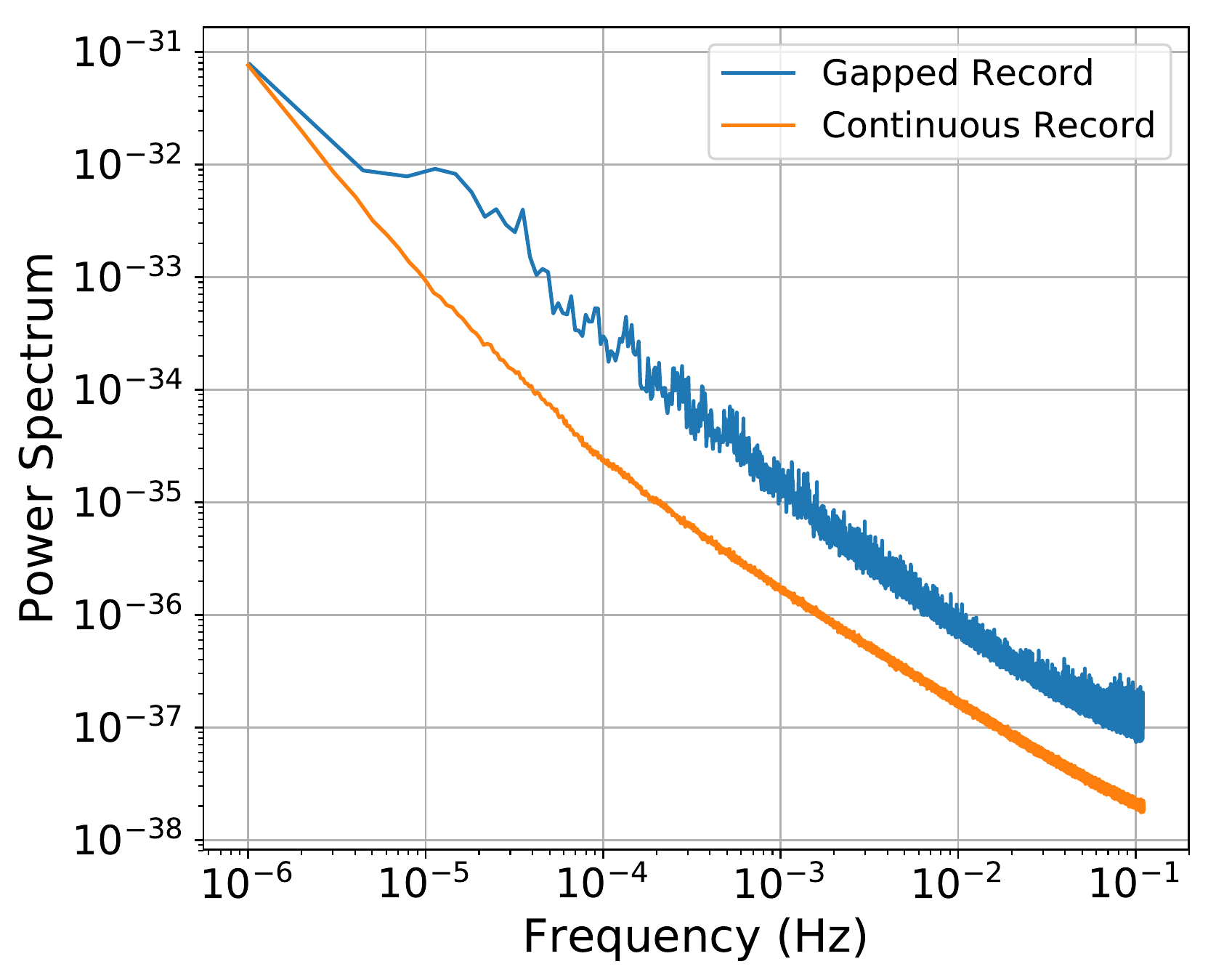}
\caption{Effect of Realistic Data Gaps. Average value of the power spectrum of 1000 independent simulations with data gaps identical to those realized in the $f_{Sr}/f_{c}$ data (blue) compared to the power spectrum of the same simulated data set without gaps (orange).}
\label{figure_s4}
\end{figure}

Finally we note that this also highlights the importance of accounting for the realistic structure of the frequency comparisons dead time in computing the expected distribution of powers in each frequency bin given the noise model of each ratio, as discussed in the main text. The pattern of observation time versus dead time in the real experiment is not as regular or periodic as what is simulated in Fig. \ref{figure_s3}, and adds additional structure to the series of peaks shown in Fig. \ref{figure_s3}(b). This is shown directly in Fig. \ref{figure_s4} where the average power spectrum of 1000 simulated data runs is plotted in comparison to the power spectrum of the data with the exact pattern of data gaps as observed in the experiment. Comparison of Fig. \ref{figure_s3}(b) and Fig. \ref{figure_s4} reveals similar general structure -- the pronounced peak at the timescale of one day and the marginally suppressed sensitivity at longer times -- but the uneven spacing of gaps and the lower total uptime of the record in Fig. \ref{figure_s4} leads to an overall higher average estimation of the power spectrum. The distribution from which the average values in Fig. \ref{figure_s4} are calculated are used to compute the corresponding 95\% confidence limit which is used for the $d_e$ limit in the main paper. The peak at the timescale of one day, $1.16\times10^{-6}$ Hz, is clearly visible at the corresponding mass of $4\times10^{-20}$ eV along with much of the structure in the limit at higher frequencies.

\subsection{Artificial Signal Injection}

\begin{figure}[ht]
\centering
\includegraphics[width=1.0\columnwidth]{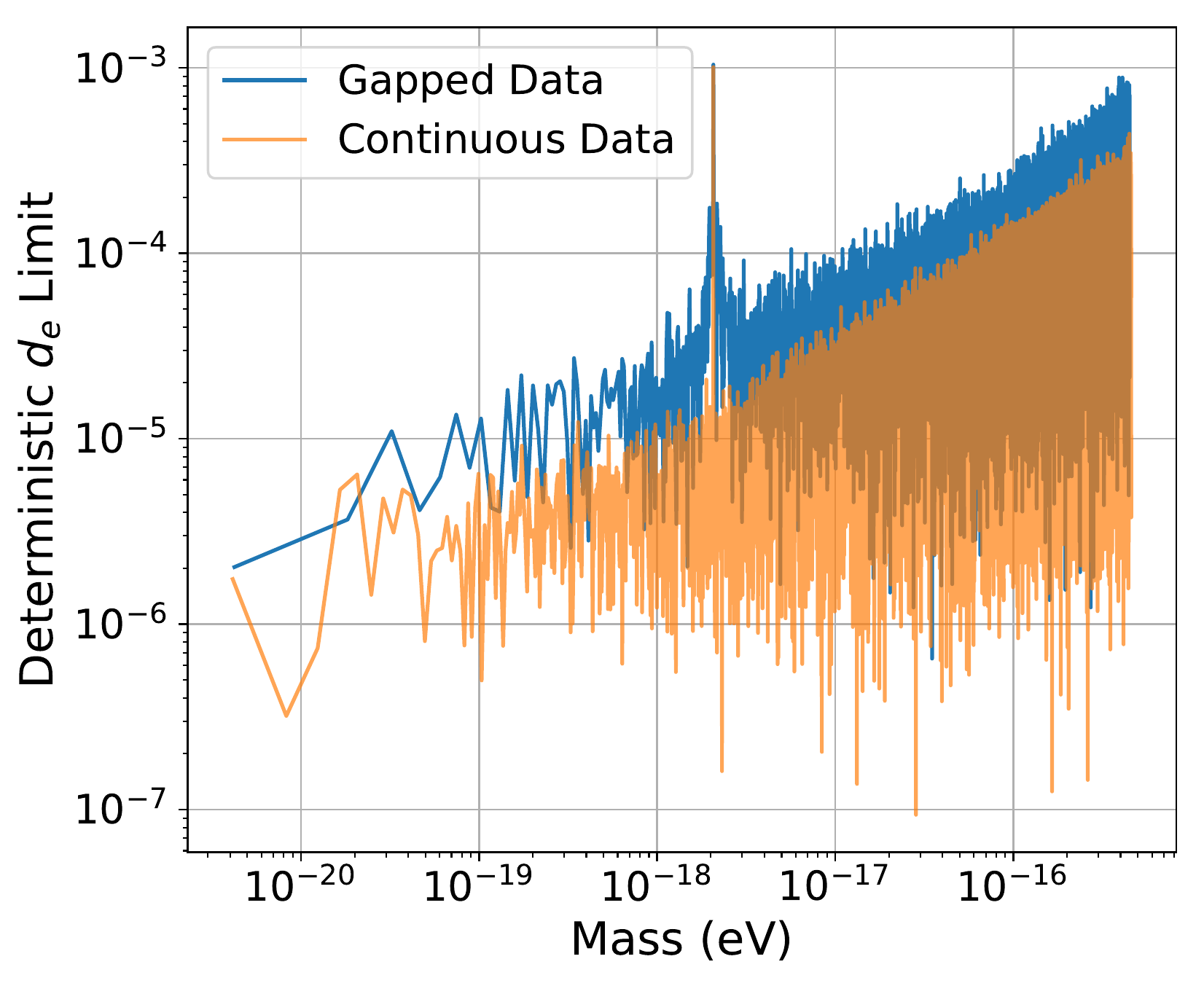}
\caption{Signal injection to the analysis pipeline. Exclusion plot for a realistic laser noise spectrum and an artificially injected periodic signal mimicking that expected from the dilatonic model of ultralight dark matter. The injected signal assumes $d_e$ of $1\times10^{-3}$, an amplitude of 1, a mass of $2\times10^{-18}$ eV, and a random phase chosen from a uniform distribution. The pattern of gaps in the gapped simulation correspond to the same gaps as the observation run in the main text.}
\label{figure_s5}
\end{figure}

Fig.~\ref{figure_s5} shows the verification of the analysis pipeline by injection of an artificial signal with known phase, frequency, and amplitude for a given value of $d_e$. Successful retrieval of the injected amplitude and modulus signals the proper normalization of both the Lomb-Scargle periodogram and the conversion from power spectra to exclusion plots. This technique is used to verify the limits for all plots in the main text.

\subsection{Maser Model and Microwave Link Noise}

\begin{figure}[ht]
\centering
\includegraphics[width=1.0\columnwidth]{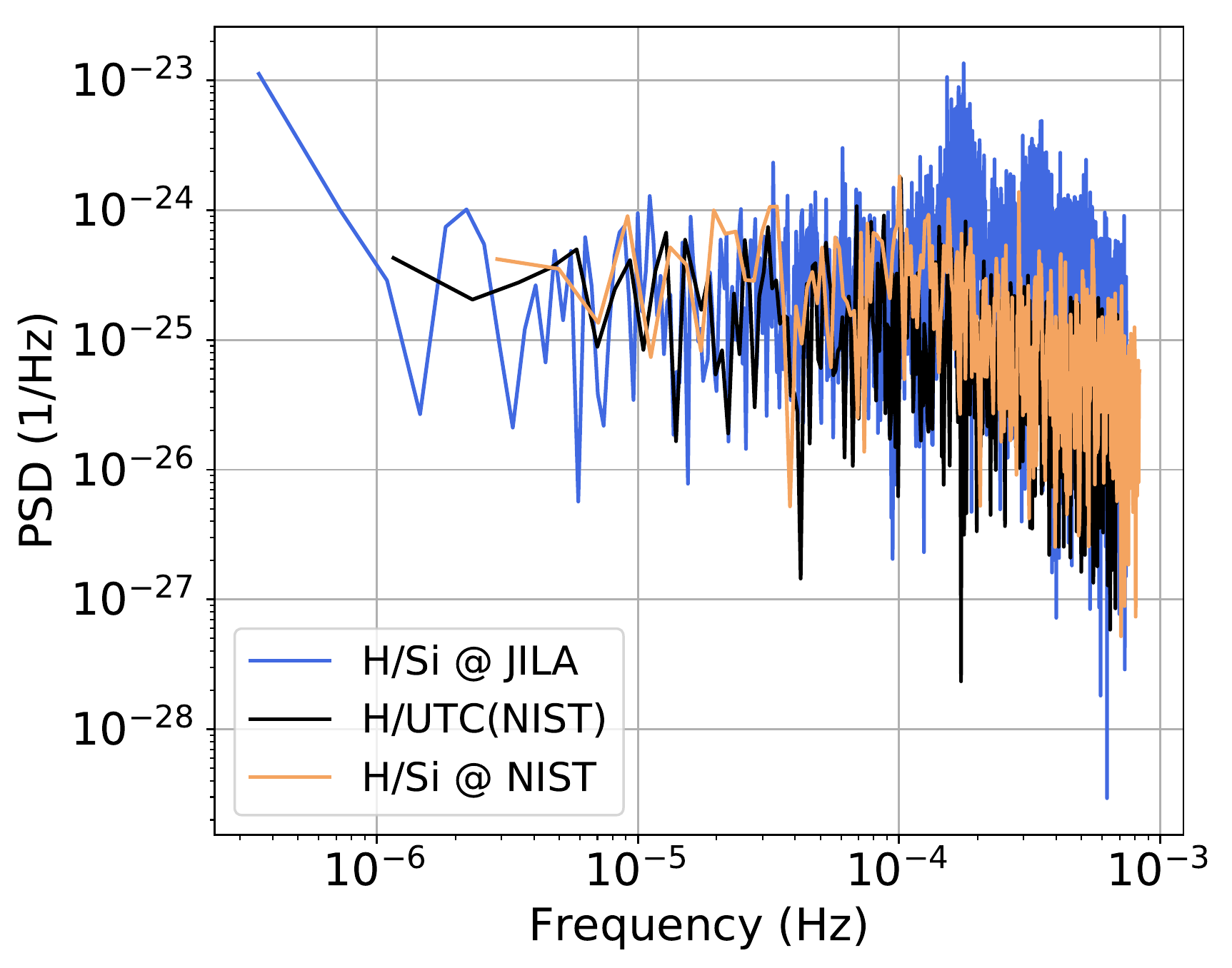}
\caption{Maser model and link noise contribution. Power spectral density of fractional fluctuations of the $f_{H}/f_{c}$ ratio as measured at JILA (blue) compared to NIST (orange). The absence of a peak at $2\times10^{-4}$ Hz in the NIST comparison shows this peak arises due to link noise. The maser comparison with UTC (black) further reinforces this conclusion.}
\label{figure_s6}
\end{figure}

Fig.~\ref{figure_s6} shows the noise added to the maser signal (ST14) by the link between JILA and NIST where the microwave signal is transferred as a modulation on an optical carrier. The noise is determined by measuring the maser signal against universal coordinated time (UTC) and comparing that to the silicon cavity (Si3) as measured at NIST versus the silicon cavity as measured at JILA. Since the apparent noise peaks appear only after the maser timing signal has been transmitted to JILA, we concluded that microwave signal transfer added this noise. We note that the direct optical frequency/phase transfer over a stabilized fiber link does not suffer from this additional noise.

\end{document}